\documentclass[preprint,
preprintnumbers,amsmath,amssymb,nofootinbib,eqsecnum]{revtex4}
\usepackage{hyperref}
\usepackage[mathscr]{eucal}
\usepackage{graphicx,wrapfig}

\begin{document}

\title{Maximal Abelian gauge and a generalized BRST transformation}

\author{Shinichi Deguchi} \email{ deguchi@phys.cst.nihon-u.ac.jp}
\affiliation{Institute of Quantum Science, College of Science and Technology, 
Nihon University, Chiyoda-ku, Tokyo 101-8308, JAPAN}
\author{Vipul Kumar Pandey}
\affiliation { Department of Physics, 
Banaras Hindu University, 
Varanasi-221005, INDIA  }
\author{Bhabani Prasad Mandal} \email { bhabani.mandal@gmail.com} 
\affiliation { Department of Physics, 
Banaras Hindu University, 
Varanasi-221005, INDIA }

\begin{abstract}
We apply a generalized Becchi-Rouet-Stora-Tyutin (BRST) formulation to establish a connection 
between the gauge-fixed $SU(2)$ Yang-Mills (YM) theories formulated in the Lorenz gauge and in the Maximal Abelian (MA) gauge. 
It is shown that the generating functional corresponding to the Faddeev-Popov (FP) effective action in the MA gauge 
can be obtained from that in the Lorenz gauge by carrying out an appropriate finite and field-dependent 
BRST (FFBRST) transformation. In this procedure, the FP effective action in the MA gauge is found 
from that in the Lorenz gauge by incorporating the contribution of non-trivial Jacobian due to the FFBRST transformation 
of the path integral measure. 
The present FFBRST formulation might be useful to see how Abelian dominance in the MA gauge is realized in the Lorenz gauge. 

\end{abstract}

\maketitle
\section{Introduction}
In the high energy region, Yang-Mills (YM) theory enjoys the asymptotic freedom 
and can be used perturbatively to describe physical systems \cite{GroWil, Politzer}. 
However, in the low energy region (or infrared region), 
coupling grows stronger and one needs to treat the theory non-perturbatively. 
Important features of YM theory at the infrared region can emerge by extracting the relevant Abelian degrees of freedom 
through the maximal Abelian (MA) projection of YM theory 
\cite{tHooft, KSW, KLSW, Kondo1, Reinhardt, Kondo-Shinohara1, Freire, IchSug1, IchSug2, KKSS}.  
The MA projection is actually performed with a partial gauge fixing called the MA gauge.  
\cite{tHooft, KSW, KLSW, Kondo1, Reinhardt, Kondo-Shinohara1, Freire, IchSug1, IchSug2, FLSS, Kondo2, SIK, KKSS}.

In $SU(N)$ YM theory, 
the MA gauge has been exploited to investigate its non-perturbative features, 
such as quark confinement \cite{Wilson}.  The MA gauge 
is a nonlinear gauge for a partial gauge fixing imposed to maintain only the maximal Abelian gauge symmetry 
specified by $U(1)^{N-1}$. This gauge enables us to extract 
Abelian degrees of freedom latent in $SU(N)$ YM theory. 
In fact, in the MA gauge, Abelian dominance \cite{EzaIwa, IchSug1, AmeSug, GIS, SugSak} 
and the emergence of magnetic monopoles \cite{tHooft, KSW, KLSW, IchSug2} 
are realized as remarkable phenomena in the non-perturbative infrared region. 
Abelian dominance is known as a low energy phenomenon in which 
only the diagonal YM fields associated with $U(1)^{N-1}$ dominate, behaving as Abelian gauge fields, 
while effects of the off-diagonal YM fields associated with $SU(N)/U(1)^{N-1}$ are strongly suppressed because 
of their large effective mass of about 1GeV \cite{AmeSug, GIS, SugSak}. 
(If we consider massive off-diagonal YM fields at the classical Lagrangian level, 
the MA gauge condition can be derived as the Euler-Lagrange equation for an additional scalar field \cite{DegKok}.) 
Magnetic monopoles emerge as topological objects characterized by the nontrivial homotopy group 
$\pi_{2} \big(SU(N)/U(1)^{N-1} \big)=\Bbb{Z}^{N-1}$ \cite{KSW}. 
The resulting effective Abelian gauge theory leads to the dual-superconductor picture for
the YM vacuum upon assuming condensation of the monopoles \cite{Nambu, tHooft2, Mandelstam}. 
In this picture, the electric flux defined from the Abelian gauge fields 
is squeezed into a string-like tube owing to the dual Meissner effect; 
as a result, (anti-)quarks are confined 
by a linear potential due to the electric flux tube \cite{AntEbe, DegKok2}. 
In this way, quark confinement is well explicated in $SU(N)$ YM theory formulated in the MA gauge.

However, since quark confinement is a physical phenomenon, 
it should be explicated independent of choices of gauge. 
We therefore need to explore how quark confinement is analytically demonstrated in terms of another gauge, for instance,  
the Lorenz gauge \cite{Lorenz}. For this purpose, it will be useful to clarify the connection between 
different gauge-fixed $SU(N)$ YM theories formulated in the MA gauge and another gauge. 
If such a connection is established, it may become possible to see how Abelian dominance 
and the emergence of magnetic monopoles are realized in another gauge. 
A universal formulation for connecting two different effective gauge theories has been developed 
by Joglekar and Mandal by means of the finite field dependent Becchi-Rouet-Stora-Tyutin (FFBRST) 
transformation \cite{sdj}.  
In this formulation, the usual (infinitesimal) BRST transformation \cite{brs, tyu} is generalized by 
allowing the parameter finite and field-dependent \cite{sdj}. 
The FFBRST transformation enjoys the properties of the usual BRST transformation except 
it does not leave the path integral measure invariant due to its finiteness. 
Under a certain condition, the non-trivial Jacobian caused by the FFBRST transformation of the path integral measure 
is expressed as a local functional of fields, which eventually modifies the effective action of the theory \cite{sdj}.
Due to this remarkable feature, the FFBRST transformation is capable of relating the generating functionals in different
gauge-fixed YM theories. 
The FFBRST formulation has found various applications in gauge field theories over last two decades 
\cite{sdj,  sdj1, rb, susk, ffanti, bss, fs, sudhak}.

In this paper, we apply the FFBRST formulation to establish a connection between the generating functional 
corresponding to the Faddeev-Popov (FP) effective action in the Lorenz gauge and that in the MA gauge.\footnote{In this paper, 
the FP effective action means the sum of the pure YM action and 
the gauge-fixing and FP ghost term that can be written in the BRST and anti-BRST exact form.}  
For this purpose, we start with the FP effective action in the Lorenz gauge \cite{BonTon, BauThi, DelJar} 
and construct the FFBRST transformation with an appropriate finite field dependent parameter. 
Then we show that the generating functional corresponding to the FP effective action in the MA gauge \cite{Kondo2, SIK, KKSS} 
can be derived from that in the Lorenz gauge by carrying out the FFBRST transformation. 
In this process, we see that the FP effective action in the MA gauge is obtained 
by incorporating a non-trivial contribution of the Jacobian arising from the FFBRST transformation of the path integral measure. 
For convenience, we treat the case of $N=2$ only. However, our approach can be generalized for arbitrary $N$.

This paper is organized as follows. 
In the next section, we briefly discuss the BRST and anti-BRST symmetries of $SU(2)$ YM theory and construct the FP effective actions both 
in the Lorenz and MA gauges. 
Sec. III is devoted to outline the FFBRST formulation in order to use it in Sec. IV. 
We present the main result of this manuscript in Sec. IV, where the connection between the generating functionals in 
the Lorenz and MA gauges is established.  Summary and concluding remarks are provided in Sec. V.

\section{BRST and anti-BRST invariant FP effective actions } 

In this section, we mention the BRST and anti-BRST transformations in $SU(2)$ YM theory 
and present the FP effective actions constructed in the MA gauge as well as in the Lorenz gauge. 
The BRST and anit-BRST invariance of the FP effective actions is ensured.

Let $A_{\mu}^{a}(x)$ $(a=1, 2, 3)$ be $SU(2)$ YM fields  
on Minkowski space with Lorentzian coordinates $(x^{\mu})$.  
The signature convention of the Minkowski metric is $(+, -, -, -)$. 
The pure YM action for $A_{\mu}^{a}$ is given by 
\begin{align}
S_\mathrm{YM} =\int d^4 x \bigg[ -\frac{1}{4} F_{\mu\nu}^{a} F^{\mu\nu a} \bigg] 
\label{2.1}
\end{align}
with the field strength 
\begin{align}
F_{\mu\nu}^{a}:=\partial_{\mu} A_{\nu}^{a} -\partial_{\nu} A_{\mu}^{a}
-g\epsilon^{abc} A_{\mu}^{b} A_{\nu}^{c} \,. 
\label{2.2}
\end{align}
Here, $g$ is a coupling constant. 
The action $S_\mathrm{YM}$ remains invariant under 
the infinitesimal gauge transformation 
\begin{align}
\delta A_{\mu}^{a} = D_{\mu} \lambda^{a} 
:=\partial_{\mu} \lambda^{a} -g\epsilon^{abc} A_{\mu}^{b} \lambda^{c} , 
\label{2.3}
\end{align}
where $\lambda^{a}$ $(a=1, 2, 3)$ are infinitesimal real functions and 
$\epsilon^{abc}$ is the Levi-Civita symbol in 3-dimensions.

We can decompose the gauge transformation (\ref{2.3}) into the $SU(2)/U(1)$ part  
specified by $\lambda^{i}$ $(i=1, 2)$ and the $U(1)$ part specified by $\lambda^{3}$ 
in such a way that  
\begin{align}
\delta A_{\mu}^{a}=\delta_{\ast} A_{\mu}^{a}+\delta_{3} A_{\mu}^{a}\,,  
\label{2.4}
\end{align}
where 
\begin{subequations}
\label{2.5}
\begin{align}
\delta_{\ast} A_{\mu}^{i} &= \nabla_{\mu} \lambda^{i} 
:=\partial_{\mu} \lambda^{i} +g\epsilon^{ij} A_{\mu}^{3} \lambda^{j} , 
\label{2.5a}
\\
\delta_{\ast} A_{\mu}^{3} &=-g\epsilon^{ij} A_{\mu}^{i} \lambda^{j} 
\qquad  \left( \epsilon^{ij}:=\epsilon^{ij3} \:\!\right) , 
\label{2.5b}
\end{align}
\end{subequations}
and 
\begin{subequations}
\label{2.6}
\begin{align}
\delta_{3} A_{\mu}^{i} &=-g\epsilon^{ij} A_{\mu}^{j} \lambda^{3} , 
\label{2.6a}
\\
\delta_{3} A_{\mu}^{3} &=\partial_{\mu} \lambda^{3} . 
\label{2.6b}
\end{align}
\end{subequations}
We see that $\nabla_{\mu}$ is the covariant derivative for 
the $U(1)$ gauge transformation (\ref{2.6}). 
The fields $A_{\mu}^{i}$ are identified as the off-diagonal YM fields 
and $A_{\mu}^{3}$ is identified as the diagonal YM field.

Next, introducing the FP ghost fields $c^{a}(x)$, 
the FP anti-ghost fields $\bar{c}^{a}(x)$, and the Nakanishi-Lautrup (NL) fields $B^{a}(x)$, 
we define the BRST transformation \cite{brs,tyu}
\begin{subequations}
\label{2.7}
\begin{align}
{s}  A_{\mu}^{a} &= -D_{\mu} c^{a} , 
\label{2.7a}
\\
{s}  c^{a} &=-\frac{1}{2}g \epsilon^{abc} c^{b} c^{c} ,
\label{2.7b}
\\
{s}  \bar{c}{}^{a} &=B^{a} ,
\label{2.7c}
\\
{s}  B^{a} &=0 \,, 
\label{2.7d}
\end{align}
\end{subequations}
and the anti-BRST transformation 
\begin{subequations}
\label{2.8}
\begin{align}
\bar{{s} } A_{\mu}^{a} &= -D_{\mu} \bar{c}{}^{a} ,  
\label{2.8a}
\\
\bar{{s} } c^{a} &=-B^{a}-g\epsilon^{abc} c^{b} \bar{c}{}^{c} . 
\label{2.8b}
\\
\bar{{s} } \bar{c}^{a} &=-\frac{1}{2}g \epsilon^{abc} \bar{c}{}^{b} \bar{c}{}^{c} ,
\label{2.8c}
\\
\bar{{s} } B^{a} &=g\epsilon^{abc} B^{b} \bar{c}{}^{c} .  
\label{2.8d}
\end{align}
\end{subequations}
Equations (\ref{2.7a}) and (\ref{2.8a}) correspond to Eq. (\ref{2.3}). 
The (anti-)BRST transformations (\ref{2.7}) and (\ref{2.8}) satisfy the nilpotency and anticommutativity: 
\begin{align}
{s} {}^2 =\bar{{s} }{}^{2}
={s}  \bar{{s} } 
+\bar{{s} } {s}  =0 \,. 
\label{2.9}
\end{align}

The (anti-)BRST transformations with a constant Grassmann parameter 
$\delta\Lambda$ are defined by 
$\delta_{\mathrm{B}}:=\delta\Lambda {s} $ and 
$\bar{\delta}_{\mathrm{B}}:=\delta\Lambda \bar{{s} }$. 
Then Eq. (\ref{2.7}) is expressed as 
\begin{subequations}
\label{2.10}
\begin{align}
\delta_{\mathrm{B}} A_{\mu}^{a} &= -\delta\Lambda \:\! D_{\mu} c^{a}  \,, 
\label{2.10a}
\\
\delta_{\mathrm{B}} c^{a} &=-\frac{1}{2} \delta\Lambda \:\! g \epsilon^{abc} c^{b} c^{c} \,,
\label{2.10b}
\\
\delta_{\mathrm{B}} \bar{c}^{\:\!a} &=\delta\Lambda \:\! B^{a} \,,
\label{2.10c}
\\
\delta_{\mathrm{B}} B^{a} &=0 \,, 
\label{2.10d}
\end{align}
\end{subequations}
and Eq. (\ref{2.8}) is expressed as 
\begin{subequations}
\label{2.11}
\begin{align}
\bar{\delta}_{\mathrm{B}} A_{\mu}^{a} &= -\delta\Lambda \:\! D_{\mu} \bar{c}{}^{a}  \,,  
\label{2.11a}
\\
\bar{\delta}_{\mathrm{B}} c^{a} &=\delta\Lambda \big( -B^{a}-g\epsilon^{abc} c^{b} \bar{c}{}^{c} \big) \,. 
\label{2.11b}
\\
\bar{\delta}_{\mathrm{B}} \bar{c}^{a} &=-\frac{1}{2} \delta\Lambda \:\! g \epsilon^{abc} \bar{c}{}^{b} \bar{c}{}^{c} \,,
\label{2.11c}
\\
\bar{\delta}_{\mathrm{B}} B^{a} &=\delta\Lambda \:\! g\epsilon^{abc} B^{b} \bar{c}{}^{c} \,.  
\label{2.11d}
\end{align}
\end{subequations}

Now we proceed to present the FP effective actions in the MA gauge as well as in the Lorenz gauge.
These actions are invariant under the BRST and anti-BRST transformations given in Eqs. (\ref{2.10}) and (\ref{2.11}).

\subsection{Lorenz gauge}

The Lorenz gauge condition $\partial^{\mu} A_{\mu}^{a}=0$ \cite{Lorenz} 
can be used to {\em completely} break the $SU(2)$ gauge invariance of the YM action (\ref{2.1}). 
This gauge condition can be incorporated into the following gauge-fixing and FP ghost term  
in a BRST and anti-BRST invariant manner \cite{BonTon, BauThi, DelJar}: 
\begin{align}
S_\mathrm{L}=\int d^4 x \bigg [ -{s}  \bar{{s} } 
\bigg( \:\! \frac{1}{2} A_{\mu}^{a} A^{\mu a} +\frac {\alpha}{2} c^{a} \bar{c}{}^{a} \bigg) \bigg] , 
\label{3.1}
\end{align}
where $\alpha$ is a gauge fixing parameter. 
Applying Eqs. (\ref{2.7}) and (\ref{2.8}) to Eq. (\ref{3.1}) and carrying out integration by parts, we obtain 
\begin{subequations}
\label{3.2}
\begin{align}
S_\mathrm{L}
&=\int d^4 x \Big[ -B^{a} \partial^{\mu} A_{\mu}^{a}
+\frac{\alpha}{2} B^{a} B^{a} 
+ \frac{\alpha}{2} g\epsilon^{abc} B^{a} c^{b} \bar{c}{}^{c} 
+\bar{c}{}^{a} {s}  (\partial^{\mu}  A_{\mu}^{a})  
-\frac{\alpha}{8} g^2 \epsilon^{abc}  \epsilon^{ade} \bar{c}{}^{b} \bar{c}{}^{c} c^{d} c^{e} \Big]  
\label{3.2a}
\\
&=\int d^4 x \Big[ -B^{a} \partial^{\mu} A_{\mu}^{a}
+\frac{\alpha}{2} B^{a} B^{a} 
+ \frac{\alpha}{2} g\epsilon^{abc} B^{a} c^{b} \bar{c}{}^{c} 
-\bar{c}{}^{a} \partial^{\mu} D_{\mu} c^{a} 
-\frac{\alpha}{8} g^2 \epsilon^{abc} \epsilon^{ade} \bar{c}{}^{b} \bar{c}{}^{c}  c^{d} c^{e} \Big] . 
\label{3.2b}
\end{align}
\end{subequations}
It is evident from Eq. (\ref{2.9}) that $\delta_{\mathrm{B}} S_\mathrm{L} =\bar{\delta}_{\mathrm{B}} S_\mathrm{L} =0$.  
Variation of $S_\mathrm{L}$ with respect to $B^{a}$ yields 
a generalized Lorenz gauge condition   
\begin{align}
\partial^{\mu} A_{\mu}^{a} -\alpha B^{a} -\frac{\alpha}{2} g\epsilon^{abc} c^{b} \bar{c}{}^{c} =0 \,.
\label{3.3}
\end{align}
(Another generalized Lorenz gauge condition $\partial^{\mu} A_{\mu}^{a} -\alpha B^{a} =0$ 
is often adopted in literature.) 
When $\alpha=0$, the gauge condition (\ref{3.3}) reduces to the (original) Lorenz gauge condition. 
The FP effective action in the Lorenz gauge is given by 
\begin{align}
\mathcal{S}_\mathrm{L}=S_\mathrm{YM}+S_\mathrm{L} \,, 
\label{EAL}
\end{align}
which is, of course, invariant under the BRST and anti-BRST transformations.

\subsection{MA gauge} 

The MA gauge condition is a nonlinear gauge condition and is defined by \cite{tHooft} 
\begin{align}
\nabla^{\mu} A_{\mu}^{i} \equiv \partial^{\mu} A_{\mu}^{i} +g\epsilon^{ij} A^{\mu 3} A_{\mu}^{j}=0 \,.  
\label{3.4}
\end{align}
This condition {\em partially} breakes the $SU(2)$ gauge invariance of the YM action (\ref{2.1}) 
so as to be maintaining its gauge invariance under the $U(1)$ gauge transformation (\ref{2.6}). 
In fact, under the gauge transformation (\ref{2.6}), $\nabla^{\mu} A_{\mu}^{i}$ transforms covariantly as
\begin{align}
\delta_{3} \big(\nabla^{\mu} A_{\mu}^{i} \big)
=-g \epsilon^{ij} \big(\nabla^{\mu} A_{\mu}^{j} \big) \lambda^{3} , 
\label{3.5}
\end{align}
so that the $U(1)$ gauge invariance is not broken. 
The MA gauge condition (\ref{3.4}) can be incorporated into 
the following gauge fixing term in a BRST and anti-BRST invariant manner \cite{Kondo2, SIK, KKSS}:  
\begin{align}
S_\mathrm{MA}=\int d^4 x \bigg[ -{s}  \bar{{s} } 
\bigg( \:\! \frac{1}{2} A_{\mu}^{i} A^{\mu i} +\frac {\beta}{2} c^{i} \bar{c}{}^{i} \bigg) \bigg] , 
\label{3.6}
\end{align}
where $\beta$ is a gauge fixing parameter. 
Applying Eqs. (\ref{2.7}) and (\ref{2.8}) into Eq. (\ref{3.6}) and carrying out integration by parts, we obtain 
\begin{subequations}
\label{3.7}
\begin{align}
S_\mathrm{MA}
&=\int d^4 x \bigg[ -B^{i} \nabla^{\mu} A_{\mu}^{i}  +\frac{\beta}{2} B^{i} B^{i} 
+\beta g\epsilon^{ij} B^{i} \bar{c}{}^{j} c^{3} 
+\bar{c}{}^{i} {s} (\nabla^{\mu}  A_{\mu}^{i})
-\frac{\beta}{4} g^2 \epsilon^{ij} \epsilon^{kl} \bar{c}{}^{i} \bar{c}{}^{j} c^{k} c^{l} \bigg] 
\label{3.7a}
\\
&=\int d^4 x \bigg[ -B^{i} \nabla^{\mu} A_{\mu}^{i}  +\frac{\beta}{2} B^{i} B^{i} 
+\beta g\epsilon^{ij} B^{i} \bar{c}{}^{j} c^{3} 
-\frac{\beta}{4} g^2 \epsilon^{ij} \epsilon^{kl} \bar{c}{}^{i} \bar{c}{}^{j} c^{k} c^{l} 
\notag 
\\
& \qquad \qquad \;\;  
-\bar{c}^{i} \big\{  \nabla^{\mu}  \nabla_{\mu} c^{i} 
-g \epsilon^{ij} \big(\nabla^{\mu}  A_{\mu}^{j} \big) c^{3} 
-g^{2} \big(A_{\mu}^{i} A^{\mu j} -\delta^{ij} A_{\mu}^{k} A^{\mu k} \big) c^{j} \big\} \bigg] . 
\label{3.7b}
\end{align}
\end{subequations}
It is easy to show that $\delta_{\mathrm{B}} S_\mathrm{MA} =\bar{\delta}_{\mathrm{B}} S_\mathrm{MA} =0$.  
Variation of $S_\mathrm{MA}$ with respect to $B^{i}$ yields a generalized MA gauge condition 
\begin{align}
\nabla^{\mu} A_{\mu}^{i} -\beta B^{i} -\beta g\epsilon^{ij} \bar{c}{}^{j} c^{3} =0 \,.
\label{3.8}
\end{align}
When $\beta=0$, this condition reduces to the (original) MA gauge condition (\ref{3.4}). 
The FP effective action in the MA gauge is given by 
\begin{align}
\mathcal{S}_\mathrm{MA}=S_\mathrm{YM}+S_\mathrm{MA} \,, 
\label{EAMA}
\end{align}
which is obviously both BRST and anti-BRST invariant.

In the light of Eq. (\ref{2.6a}), we can consistently impose the $U(1)$ gauge transformation rules 
\begin{subequations}
\label{3.9}
\begin{align}
\delta_{3} B^{i} &=-g\epsilon^{ij} B^{j} \lambda^{3} , 
\label{3.9a}
\\
\delta_{3} c^{i} &=-g\epsilon^{ij} c^{j} \lambda^{3} , 
\label{3.9b}
\\
\delta_{3} \bar{c}^{i} &=-g\epsilon^{ij} \bar{c}^{j} \lambda^{3} 
\label{3.9c}
\end{align}
\end{subequations}
on the fields $B^{i}$, $c^{i}$, and $\bar{c}^{i}$. 
Then it is clear that $A_{\mu}^{i} A^{\mu i}$ and $c^{i} \bar{c}{}^{i}$ remain invariant 
under the $U(1)$ gauge transformation. 
Consequently, it follows from Eq. (\ref{3.6}) that the gauge-fixing and FP ghost term $S_\mathrm{MA}$ 
is $U(1)$ gauge invariant.

\section{Outline of the FFBRST formulation}

In this section, we recapitulate the FFBRST formulation for YM theory developed in Ref. \cite{sdj}. 
For this purpose, we first write the usual BRST transformation (\ref{2.10}) as  
\begin{equation}
\delta_\mathrm{B} \phi_I (x) =\delta \Lambda s\phi_I (x) , 
\label{4.1}
\end{equation} 
where $\delta\Lambda$ is an infinitesimal and field-independent 
Grassmann parameter,\footnote{Here the ``infinitesimal" means that $\delta\Lambda$ can be expressed as $\delta\Lambda=\Lambda dk$ 
using some Grassmann parameter $\Lambda$ and an infinitesimal commuting real number $dk$. 
In this sense, $\Lambda$ is naively treated as ``finite".}    
and $\phi_I$ is the generic notation of the fields $(A^{a}_{\mu}, c^{a}, \bar{c}^{a}, B^{a})$ involved in the theory.  
The index $I$ distinguishes the fields as well as their components. 
The basic properties of BRST transformation  do not depend on whether 
the parameter $\delta\Lambda$  is (i) finite or infinitesimal and/or (ii) field-dependent or not, as long 
as it is anti-commuting and spacetime independent. This renders
us a freedom to construct the BRST transformation with the parameter finite and field-dependent without 
affecting its basic features.  First we make the infinitesimal parameter field-dependent 
by interpolating a continuous parameter, $\kappa\ (0\leq \kappa\leq 1)$, in the theory.
The generic field, $\phi_I (x,\kappa)$, depends on $\kappa$ such that $\phi_I (x,\kappa =0)=\phi_I (x)$ is the initial field and 
$\phi_I (x,\kappa =1)=\phi_I^\prime(x)$ is the transformed field.

The infinitesimal field-dependent BRST transformation is now defined by \cite{sdj} 
\begin{align} 
d \phi_I (x, \kappa) =\Theta^\prime [\phi (\kappa ) ] s\phi_I (x,k) d\kappa \,, 
\label{diff}
\end{align}
in accordance with Eq. (\ref{4.1}). 
Here, $\Theta^\prime [\phi ( \kappa ) ]{d\kappa}$ is an infinitesimal but field-dependent Grassmann parameter. 
The FFBRST transformation, $\phi_I (x) \rightarrow \phi_I^\prime(x)$, is then provided 
by integrating the infinitesimal transformation (\ref{diff}) from $\kappa =0$ to $\kappa= 1$, as follows: 
\begin{equation}
\phi_{I}^\prime(x) \equiv \phi_{I} (x,\kappa =1)=\phi_{I} (x,\kappa=0)+\Theta[\phi] s\phi_{I} (x) \,, 
\label{kdep}
\end{equation}
where 
\begin{equation}
\Theta [\phi] 
=\int_{0}^{1} 
\Theta^\prime [\phi (\kappa ) ] d\kappa 
= \Theta ^\prime [\phi] \frac{ \exp f[\phi]-1}{f[\phi]},
\label{80}
\end{equation}
is the finite field-dependent parameter and $f[\phi]$ is given by\cite{sdj} 
\begin{eqnarray}
f[\phi]= \sum_I \int d^4x\:\! s \phi_I (x) \frac{ \delta \Theta ^\prime [\phi]}{\delta\phi_I(x)} \,.
\label{f}
\end{eqnarray}  
In Eqs. (\ref{80}) and (\ref{f}), $\phi$ can be understood as $\phi(\kappa=0)$. 
The resulting FFBRST transformation in Eq. (\ref{kdep}) leaves the FP effective action of 
the theory invariant but the 
functional integral changes non-trivially under it due to the presence of finite parameter \cite{sdj}.
Now we briefly outline how to compute the Jacobian of path integral measure for the FFBRST transformation.

The path integral measure $\mathcal{D}\phi :=\prod_{I} \prod_{x} d\phi_{I}(x)$ transforms 
under the FFBRST transformation 
$\phi_{I}(x) \rightarrow \phi_{I} (x,\kappa)=\phi_{I}(x)+\Theta[\phi, \kappa] s\phi_{I} (x)$, with 
$\Theta [\phi, \kappa] :=\int_{0}^{\kappa} \Theta^\prime [\phi (\tilde{\kappa}) ] d\tilde{\kappa}^{\!\:}$,  
according to  
\begin{eqnarray}
{\cal D}\phi &=&J(\kappa) {\cal D}\phi(\kappa)\,,  
\label{26}
\end{eqnarray}
where $J(\kappa)$ is the Jacobian for the present FFBRST transformation and satisfies $J(0)=1$. 
It has been shown \cite{sdj} that
the Jacobian $J(\kappa )$ can be replaced within the functional integral as 
\begin{equation}
J(\kappa ) \,\Rightarrow\, \exp\{ iS_1 [ \phi(\kappa), \kappa ] \}, 
\label{s}
\end{equation}
iff the following condition is satisfied \cite{sdj}:
\begin{eqnarray}
\int {\cal D}\phi(\kappa)   \bigg[\:\!  \frac{1}{J(\kappa)}\frac{dJ(\kappa)}{d\kappa}-i \:\! \frac
{dS_1[\phi (\kappa ), \kappa]}{d\kappa} \bigg] 
\exp \{ \:\! i(\mathcal{S}[\phi(\kappa) ]+S_1[\phi(\kappa),\kappa ]) \}=0 \,. 
\label{mcond}
\end{eqnarray}
Here, $S_1[\phi (\kappa), \kappa]$ is a local functional of the fields, and 
$\mathcal{S}$ denotes either the FP effective action $\mathcal{S}_\mathrm{L}$ or $\mathcal{S}_\mathrm{MA}$. 
The infinitesimal change in the Jacobian $J(\kappa)$
can be calculated with the following formula \cite{sdj} 
\begin{equation}
\frac{1}{J(k)}\frac{dJ(k)}{d\kappa}=-\sum_{I} \int d^4y \Bigg[ (-1)^{|I|} \frac{
\delta\Theta^\prime [\phi(\kappa) ]}{\delta\phi_{I} (y,\kappa )} \:\! s\phi_{I} (y,\kappa )\Bigg], 
\label{jac}
\end{equation}
where $|I|$ is defined as $|I|=0$ for bosonic fields $\phi_I$ and 
as $|I|=1$ for fermionic fields $\phi_I$. 
Once we know $J^{-1} (dJ/d\kappa)$, we can find $S_1$ from the
condition in Eq. (\ref{mcond}). 
 

\section{Connection between generating functionals in the Lorenz and MA gauges}
In this section, we construct the FFBRST transformation with an appropriate finite parameter to obtain
the generating functional corresponding to $\mathcal{S}_\mathrm{MA}$ from that corresponding to $\mathcal{S}_\mathrm{L}$. 
We calculate the Jacobian corresponding to such a FFBRST transformation 
following the method outlined in Sec III and show that
it is a local functional of fields and accounts for the differences of the two FP effective actions.

The generating functional corresponding to the FP effective action $\mathcal{S}_\mathrm{L}$ is written as 
\begin{equation}
Z_\mathrm{L} = \int {\cal D} \phi \exp{(i\mathcal{S}_\mathrm{L}[\phi])} \,. 
\label{GFL}
\end{equation}
Now, to obtain the generating functional corresponding $\mathcal{S}_\mathrm{MA}$, we apply the FFBRST transformation 
(\ref{kdep}) with a finite parameter $\Theta[\phi]$ obtainable according to Eq. (\ref{80})   
from the infinitesimal but field dependent parameter $\Theta' [\phi(\kappa)] d\kappa$ defined by 
\begin{align}
& \Theta' [\phi(\kappa)] 
\notag
\\
& = i\int d^4 x \big[\gamma_1 \bar{c}^i{B^i}+\gamma_2\bar{c}^3{B^3}+\gamma_3 
\big\{ \bar{c}^a(\partial^{\mu} A^{a}_{\mu})-\bar{c}^i (\nabla^{\mu} A^{i}_{\mu}) \big\} 
+\gamma_4 g \epsilon^{abc}\bar{c}^a\bar{c}^b{c^c}+\gamma_5 g \epsilon^{ij} \bar{c}^i {\bar c}^j{c^3} \big] .  
\label{para}
\end{align}
Here, $\gamma_p \ (p=1,2,3,4,5)$ are arbitrary constant parameters and all the fields depend 
on the parameter $\kappa $.  The infinitesimal change in the Jacobian corresponding to this FFBRST transformation 
is calculated using Eq. (\ref{jac}) to obtain 
\begin{align}
\frac{1}{J}\frac{dJ}{dk} &= -i\int d^4 x  \bigg[-\gamma_1 B^i B^i -\gamma_2 B^3 B^3 
+\gamma_3 \big\{\bar{c}^a s(\partial^{\mu} A^{a}_{\mu}) 
-\bar{c}^i s(\nabla^{\mu} A^{i}_{\mu}) \big\}
\notag
\\
& \qquad -\gamma_3{B^a}\partial^{\mu} A^{a}_{\mu}
+\gamma_3{B^i}\nabla^{\mu} A^{i}_{\mu} 
+\gamma_4 \bigg\{-2g\epsilon^{abc}{B^a}{\bar c}^b{c}^c+\frac{1}{2} g^2 \epsilon^{abc} \epsilon^{ade} \bar{c}^b\bar{c}^c {c^d}{c^e} \bigg\}
\notag
\\
& \qquad +\gamma_5 \bigg\{-2g\epsilon^{ij}{B^i}{\bar c}^j{c^3}+\frac{1}{2} g^2 \epsilon^{ij} \epsilon^{kl} {\bar c}^i {\bar c}^j {c^k}{c^l} \bigg\} \bigg] .  
\label{jac1}
\end{align}

To express the Jacobian contribution in terms of a local functional of fields, 
we make an {\em ansatz} for $S_1$ by considering all possible terms that could arise from such a transformation as  
\begin{align}
& S_1[\phi(\kappa), \kappa ] 
\notag
\\
&=\int d^4 x \big[\:\! \xi_1 B^a {\partial^{\mu}A^a_{\mu}}+\xi_2{B^i}{\nabla^{\mu} A^{i}_{\mu}}
+\xi_3 B^a B^a +\xi_4 B^i B^i +\xi_5\bar{c}^i s(\nabla^{\mu} A^{i}_{\mu})
+\xi_6\bar{c}^a s(\partial^{\mu} A^{a}_{\mu}) 
\nonumber \\ 
& \qquad +\xi_7 g \epsilon^{abc}{B^a}{c}^b{\bar c}^c+\xi_8 g^2\epsilon^{abc}\epsilon^{ade}\bar{c}^b\bar{c}^c{c^d}{c^e}+\xi_9 g\epsilon^{ij}{B^i}{\bar c}^j{c^3}
+\xi_{10}g^2\epsilon^{ij}\epsilon^{kl}\bar{c}^i\bar{c}^j{c^k}{c^l} \!\: \big] ,  
\label{S1}
\end{align}
where all the fields are considered to be $\kappa$ dependent and we have introduced arbitrary $\kappa$ dependent parameters 
$\xi_n =\xi_n(\kappa) \ (n=1,2\ldots, 10)$. It is straight to calculate 
\begin{align}
\frac {dS_1}{dk} &=\int d^4 x \bigg[ \:\! \xi'_1 B^a {\partial^{\mu}A^a_{\mu}}+{\xi'_2 B^i}{\nabla^{\mu} A^{i}_{\mu}}
+\xi'_3 B^a B^a +\xi'_4 B^i B^i + \xi'_5\bar{c}^i s({\nabla^{\mu} A^{i}_{\mu}})+\xi'_6 \bar{c}^a s(\partial^{\mu} A^{a}_{\mu}) 
\nonumber \\ 
& \qquad  +\xi'_7 g \epsilon^{abc}{B^a}{c}^b{\bar c}^c +\xi'_8 g^2\epsilon^{abc}\epsilon^{ade}\bar{c}^b\bar{c}^c{c^d}{c^e}+\xi'_9 g\epsilon^{ij}{B^i}{\bar c}^j{c^3}
+\xi'_{10}g^2\epsilon^{ij}\epsilon^{kl}\bar{c}^i\bar{c}^j{c^k}{c^l}
\nonumber \\ 
& \qquad  
+\Theta' \bigg\{ \xi_1{B^a} s({\partial^{\mu}A^a_{\mu}}) + \xi_2{B^i} s({\nabla^{\mu} A^{i}_{\mu}}) 
+\xi_5 B^{i} s({\nabla^{\mu} A^{i}_{\mu}})+\xi_6 {B}^a s({\partial^{\mu} A^{a}_{\mu}}) 
\nonumber \\
& \qquad 
+\frac{1}{2} (\xi_7 +4\xi_8) g^2\epsilon^{abc}\epsilon^{ade} {B}^b\bar{c}^c {c^d}{c^e}  
+\frac{1}{2} (\xi_9 +4\xi_{10}) g^2 \epsilon^{ij} \epsilon^{kl} {B^i} \bar{c}^j {c^k}{c^l} 
\bigg\} \bigg]  
\label{S1/kappa}
\end{align}
with $\xi_{n}^{\prime}:=d\xi_{n}/d\kappa$ 
by using Eqs. (\ref{diff}) and (\ref{2.7}) and the nilpotency $s^2 =0$.  
We substitute Eqs. (\ref{jac1}) and (\ref{S1/kappa}) into Eq. (\ref{mcond}) with $\mathcal{S}=\mathcal{S}_\mathrm{L}$  
to find the condition to replace the Jacobian contribution in terms of a local functional of the fields as 
\begin{align} 
&\int D\phi(\kappa)  \exp [ \:\! i(\mathcal{S}_\mathrm{L} [\phi(\kappa)]+S_1[\phi(\kappa),\kappa ]) ]  
\notag
\\
& \times \int d^4 x  
\bigg[-\gamma_1 B^i B^i -\gamma_2 B^3 B^3 
+\gamma_3 \big\{\bar{c}^a s(\partial^{\mu} A^{a}_{\mu}) 
-\bar{c}^i s(\nabla^{\mu} A^{i}_{\mu}) \big\}
\notag
\\
& \qquad -\gamma_3{B^a}\partial^{\mu} A^{a}_{\mu}
+\gamma_3{B^i}\nabla^{\mu} A^{i}_{\mu} 
+\gamma_4 \bigg\{-2g\epsilon^{abc}{B^a}{\bar c}^b{c}^c+\frac{1}{2} g^2 \epsilon^{abc} \epsilon^{ade} \bar{c}^b\bar{c}^c {c^d}{c^e} \bigg\}
\notag
\\
& \qquad +\gamma_5 \bigg\{-2g\epsilon^{ij}{B^i}{\bar c}^j{c^3}+\frac{1}{2} g^2 \epsilon^{ij} \epsilon^{kl} {\bar c}^i {\bar c}^j {c^k}{c^l} \bigg\} 
\notag
\\
& \qquad +\xi'_1 B^a {\partial^{\mu}A^a_{\mu}}+{\xi'_2 B^i}{\nabla^{\mu} A^{i}_{\mu}}
+\xi'_3 B^a B^a +\xi'_4 B^i B^i + \xi'_5\bar{c}^i s({\nabla^{\mu} A^{i}_{\mu}})+\xi'_6 \bar{c}^a s(\partial^{\mu} A^{a}_{\mu}) 
\nonumber \\ 
& \qquad  +\xi'_7 g \epsilon^{abc}{B^a}{c}^b{\bar c}^c +\xi'_8 g^2\epsilon^{abc}\epsilon^{ade}\bar{c}^b\bar{c}^c{c^d}{c^e}+\xi'_9 g\epsilon^{ij}{B^i}{\bar c}^j{c^3}
+\xi'_{10}g^2\epsilon^{ij}\epsilon^{kl}\bar{c}^i\bar{c}^j{c^k}{c^l}
\nonumber \\ 
& \qquad  
+\Theta' \bigg\{ \xi_1{B^a} s({\partial^{\mu}A^a_{\mu}}) + \xi_2{B^i} s({\nabla^{\mu} A^{i}_{\mu}}) 
+\xi_5 B^{i} s({\nabla^{\mu} A^{i}_{\mu}})+\xi_6 {B}^a s({\partial^{\mu} A^{a}_{\mu}}) 
\nonumber \\
& \qquad 
+\frac{1}{2} (\xi_7 +4\xi_8) g^2\epsilon^{abc}\epsilon^{ade} {B}^b\bar{c}^c {c^d}{c^e}  
+\frac{1}{2} (\xi_9 +4\xi_{10}) g^2 \epsilon^{ij} \epsilon^{kl} {B^i} \bar{c}^j {c^k}{c^l} 
\bigg\} \bigg]  
=0 \,.  
\label{bcon}
\end{align}
This can be written as  
\begin{align} 
&\int D \phi(\kappa) \exp [ \:\! i(\mathcal{S}_\mathrm{L} [\phi(\kappa)]+S_1[\phi(\kappa),\kappa ]) ]
\notag
\\
& \times \int d^4 x  
\bigg[(-\gamma_1 +\xi'_3 +\xi'_4) B^i B^i 
+(-\gamma_2 +\xi'_3) B^3 B^3 
+(\gamma_3 +\xi'_6) \bar{c}^a s(\partial^{\mu} A^{a}_{\mu}) 
\notag
\\
& \qquad 
+(-\gamma_3 +\xi'_5) \bar{c}^i s(\nabla^{\mu} A^{i}_{\mu}) 
+(-\gamma_3 +\xi'_1) B^a \partial^{\mu} A^{a}_{\mu} 
+(\gamma_3 +\xi'_2) B^i \nabla^{\mu} A^{i}_{\mu} 
\notag
\\
& \qquad 
+(-2\gamma_4 +\xi'_7) g\big(\epsilon^{ij} B^i c^j \bar{c}^3 +\epsilon^{ij} B^3 c^i \bar{c}^j \big) 
+(-2\gamma_4 -2\gamma_5 +\xi'_7 +\xi'_9) g\epsilon^{ij} B^i \bar{c}^j c^3 
\notag
\\
& \qquad 
+\frac{1}{2} (\gamma_4 +\gamma_5 +2\xi'_8 +2\xi'_{10}) g^2 \epsilon^{ij} \epsilon^{kl} \bar{c}^i \bar{c}^j c^k c^l 
+2(\gamma_4 +2\xi'_8) g^2 \epsilon^{ij} \epsilon^{ik} \bar{c}^j \bar{c}^3 c^k c^3 
\notag
\\
& \qquad 
+\Theta' \bigg\{ 
(\xi_1+\xi_6) {B^a} s(\partial^{\mu} A^{a}_{\mu})
+(\xi_2+\xi_5) {B^i} s(\nabla^{\mu} A^{i}_{\mu}) 
\notag
\\
& \qquad 
+\frac{1}{2} ( \xi_7 +4\xi_8+ \xi_9 +4\xi_{10} ) g^2 \epsilon^{ij}\epsilon^{kl} B^i {\bar c}^j {c^k}{c^l} 
\notag
\\
& \qquad 
+(\xi_7 +4\xi_8) g^2 \big(\epsilon^{ij}\epsilon^{ik} B^j {\bar c}^3 {c^k}{c^3} 
-\epsilon^{ij}\epsilon^{ik}{B^3}{\bar c}^j {c^k}{c^3} \big) \bigg\} \bigg] =0 \,. 
\label{bcon2}
\end{align}
The terms proportional to $\Theta'$, which are regarded in Eq. (\ref{bcon2}) as nonlocal terms due to $\Theta'$, 
independently vanish if 
\begin{subequations}
\label{37}
\begin{align}
& \xi_1+\xi_6=0 \,, 
\label{37a}
\\
& \xi_2+\xi_5=0 \,, 
\label{37b}
\\
& \xi_7+4\xi_8+\xi_9+4\xi_{10}=0 \,, 
\label{37c} 
\\
&\xi_7+4\xi_8=0 \,.
\label{37d}
\end{align}
\end{subequations}
To make the remaining local terms in Eq. (\ref{bcon2}) vanish, we need the following conditions: 
\begin{subequations}
\label{38}
\begin{align}
& \xi'_1 -\gamma_3 =0 \,,
\label{38a}
\\
& \xi'_2 +\gamma_3 =0 \,, 
\label{38b} 
\\
& \xi'_3 -\gamma_2 =0 \,,
\label{38c}
\\
& \xi'_3 +\xi'_4 -\gamma_1 =0 \,,
\label{38d}
\\
& \xi'_5 -\gamma_3 =0 \,,
\label{38e}
\\
& \xi'_6 +\gamma_3 =0 \,,
\label{38f}
\\
& \xi'_7 -2\gamma_4 =0 \,, 
\label{38g}
\\
& \xi'_7 +\xi'_9 -2(\gamma_4 +\gamma_5) =0 \,, 
\label{38h}
\\
& \xi'_8 +\frac{1}{2} \gamma_4  =0 \,. 
\label{38i}
\\
& \xi'_8 +\xi'_{10} + \frac{1}{2} (\gamma_4 +\gamma_5) =0 \,,
\label{38j}
\end{align}
\end{subequations}
from which we also have
\begin{subequations}
\label{39}
\begin{align}
& \xi'_4 -\gamma_1 +\gamma_2 =0 \,,
\label{39a}
\\
& \xi'_9 -2\gamma_5 =0 \,,
\label{39b}
\\
& \xi'_{10} +\frac{1}{2} \gamma_5 =0 \,.
\label{39c}
\end{align}
\end{subequations}
The differential equations for $\xi_n (\kappa)$ can indeed be solved with the initial conditions  
$\xi_n (0)=0$ to obtain the solutions  
\begin{alignat}{5}
& \xi_1 =\gamma_3 \kappa \,, 
& \quad 
& \xi_2 =-\gamma_3 \kappa \,,
& \quad 
& \xi_3 =\gamma_2 \kappa \,, 
\notag
\\
& \xi_4 =(\gamma_1 -\gamma_2) \kappa \,, \;
& \quad 
& \xi_5 =\gamma_3 \kappa \,,
& \quad 
& \xi_6 =-\gamma_3 \kappa \,, 
\notag
\\
& \xi_7 =2\gamma_4 \kappa \,, 
& \quad 
& \xi_8 =-\frac{1}{2} \gamma_4 \kappa \,,
& \quad \;
& \xi_9 =2\gamma_5 \kappa \,, 
\notag
\\
& \xi_{10} =-\frac{1}{2} \gamma_5  \kappa \,. 
\label{40}
\end{alignat}
It should be noted that the solutions in Eq (\ref{40}) also satisfy Eqs. (\ref{37a})--(\ref{37d}). 
The conditions in Eqs. (\ref{37}) and (\ref{38}) are thus compatible with each other.

Since $\gamma_p \ (p=1,2,3,4,5)$ are arbitrary constant parameters, 
we can chose them as follows:  
\begin{alignat}{5}
& \gamma_1 = \frac{1}{2}({\beta-\alpha}) \,, 
& \quad \;
& \gamma_2 = -\frac{\alpha}{2} \,, 
& \quad \;
& \gamma_3 = 1 \,, 
\notag
\\
& \gamma_4=-\frac{\alpha}{4} \,, 
& \quad 
& \gamma_5 =\frac{\beta}{2}  \,. 
\label{41}
\end{alignat}
Substituting the solutions found in Eq. (\ref{40}) into Eq. (\ref{S1}) and considering the specific values of the parameters in Eq. (\ref{41}), 
we obtain  
\begin{align}
S_1[\phi(1), 1 \:\!] 
&=\int d^4 x \bigg[ B^a {\partial^{\mu}A^a_{\mu}} -{B^i}{\nabla^{\mu} A^{i}_{\mu}}
-\frac{\alpha}{2} B^a B^a +\frac{\beta}{2} B^i B^i +\bar{c}^i s(\nabla^{\mu} A^{i}_{\mu})
-\bar{c}^a s(\partial^{\mu} A^{a}_{\mu}) 
\notag
\\ 
& \qquad -\frac{\alpha}{2} g \epsilon^{abc}{B^a}{c}^b{\bar c}^c +\frac{\alpha}{8} g^2\epsilon^{abc}\epsilon^{ade}\bar{c}^b\bar{c}^c{c^d}{c^e}
+\beta g\epsilon^{ij}{B^i}{\bar c}^j{c^3}
-\frac{\beta}{4} g^2\epsilon^{ij}\epsilon^{kl}\bar{c}^i\bar{c}^j{c^k}{c^l}  \bigg] .   
\label{42} 
\end{align}
Thus the FFBRST transformation with the finite parameter $\Theta$ that is defined by Eq. (\ref{80}) with Eq. (\ref{para})  
changes the generating functional $Z_\mathrm{L}$ as 
\begin{align}
Z_\mathrm{L}= \int {\cal D } \phi \exp (i \mathcal{S}_\mathrm{L} [\phi]) 
\;\stackrel{\mbox{FFBRST}}{\longrightarrow} \;
&\int {\cal D } \phi^{\prime} \exp \{ i(\mathcal{S}_\mathrm{L}[\phi^{\prime}\:\!] +S_1[\phi^{\prime}, 1 \:\!]) \} 
\notag
\\
=&\int {\cal D } \phi \exp \{ i(\mathcal{S}_\mathrm{L}[\phi] +S_1[\phi, 1 \:\!]) \} 
\notag
\\
=&\int {\cal D } \phi \exp (i\mathcal{S}_\mathrm{MA}[\phi]) =Z_\mathrm{MA} \,,   
\label{43}
\end{align}
where Eqs. (\ref{3.2a}), (\ref{3.7a}), and (\ref{42}) have been used to see that 
$\mathcal{S}_\mathrm{L}[\phi] +S_1[\phi, 1^{\:\!}] =\mathcal{S}_\mathrm{MA}[\phi]$. 
In this way, the suitably constructed FFBRST transformation maps $SU(2)$ YM theory in the Lorenz gauge to 
that in the MA gauge.

\section{Summary and Concluding remarks}

We have applied the FFBRST formulation developed in Ref. \cite{sdj} to clarify the connection  
between the gauge-fixed $SU(2)$ YM theories formulated in the Lorenz and MA gauges. 
We have explicitly shown that the generating functional corresponding to the FP effective action in the MA gauge 
can be obtained from that in the Lorenz gauge by carrying out a suitably constructed FFBRST transformation (see Eq. (\ref{43})). 
In this procedure, the FP effective action in the MA gauge is found from that in the Lorenz gauge by taking into account 
the non-trivial Jacobian arising from the FFBRST transformation of the path integral measure.

In this paper, we have considered only the FFBRST transformation. 
However, since both the FP effective actions given in Eqs. (\ref{EAL}) and (\ref{EAMA}) are invariant under 
the anti-BRST transformation (\ref{2.11}), we can construct the finite field dependent anti-BRST (FF anti-BRST) transformation  
\cite{susk,ffanti,ffanti1} as a counterpart of the FFBRST transformation.

Now, it is, of course, possible to derive the generating functional corresponding to the FP effective action in the Lorenz gauge 
from that in the MA gauge by applying the inverse FFBRST transformation that is formally defined by replacing 
$\gamma_p$ in Eq. (\ref{para}) with $-\gamma_p\:\!$: 
\begin{align}
Z_\mathrm{MA} 
\;\stackrel{\mbox{inverse FFBRST}}{\longrightarrow} \; Z_\mathrm{L} \,.
\label{44}
\end{align}
As we have mentioned in Sec. I, Abelian dominance is realized in the MA gauge owing to 
the large effective mass of off-diagonal YM fields evaluated in this gauge. 
In the case of $N=2$, the Abelian dominance phenomenon can be effectively incorporated in 
the present FFBRST formulation by adding the following mass term 
to the FP effective action $\mathcal{S}_\mathrm{MA}$: 
\begin{align}
S_{m}=\int d^4 x \bigg[\;\! \frac{1}{2} m^2 A_{\mu}^{i} A^{\mu i} \bigg] , 
\label{45}
\end{align}
where $m$ denotes an effective mass of the off-diagonal YM fields $A_{\mu}^{i}$.\footnote{We can 
consider the Curci-Ferrari mass term \cite{CurFer, BanDeg} 
\begin{align*}
\widetilde{S}_{m}=\int d^4 x \bigg[\;\! \frac{1}{2} m^2 \big( A_{\mu}^{i} A^{\mu i} +2\beta c^i \bar{c}^i \big) \bigg] , 
\end{align*}
as an alternative to the simple mass term $S_{m}$. 
Remarkably, $\widetilde{S}_{m}$ is BRST and anti-BRST invariant {\em on-shell} 
in the sense that the invariance can be shown with the aid of Eq. (\ref{3.8}). 
The mass term $\widetilde{S}_{m}$ also remains invariant under the $U(1)$ gauge transformation 
specified by Eqs. (\ref{2.6a}), (\ref{3.9b}), and (\ref{3.9c}). 
Since $\widetilde{S}_{m}$ possesses the on-shell BRST and anti-BRST invariance, this term may be more convenient 
for describing Abelian dominance in the FFBRST formulation.}  
The mass term $S_{m}$ remains invariant under the $U(1)$ gauge transformation (\ref{2.6a}), so that 
it does not break the $U(1)$ gauge invariance of $\mathcal{S}_\mathrm{MA}$. 
Being introduced $S_{m}$, Eq. (\ref{44}) is modified as 
\begin{align}
&\widehat{Z}_\mathrm{MA}=\int {\cal D } \phi \exp \{i (\mathcal{S}_\mathrm{MA}[\phi]+S_{m}[A] ) \}  
\notag
\\
&\;\stackrel{\mbox{inverse FFBRST}}{\longrightarrow} \; 
\widehat{Z}_\mathrm{L}
=\int {\cal D } \phi \exp \{ i (\mathcal{S}_\mathrm{L}[\phi] +S'_{m} [\phi] ) \} , 
\label{46}
\end{align}
where $S'_{m}$ is defined as the inverse FFBRST transformation of $S_{m}$. 
As expected $S'_{m}$ is highly nonlocal and will not be easy to deal with. 
However, $S'_{m}$ must describe a phenomenon corresponding to Abelian dominance, 
and we would be able to see with $S'_{m}$ how Abelian dominance is realized in the Lorenz gauge. 
We therefore hope to investigate the details of $S'_{m}$ in the near future.

\section*{Acknowledgments}
This work is supported in part by JSPS BRIDGE Fellowship Program (No. BR150602). 
BPM acknowledges the hospitality of Institute of Quantum Science, Nihon University.

\end{document}